# The Trajectory of Romance Scams in the U.S.

LD Herrera
*The Beacom College of Computer and Cyber Sciences*
*Dakota State University*
Madison, SD, USA
ld.herrera@trojans.dsu.edu
0009-0005-8027-008X

John Hastings
*The Beacom College of Computer and Cyber Sciences*
*Dakota State University*
Madison, SD, USA
john.hastings@dsu.edu
0000-0003-0871-3622

*Abstract*—Romance scams (RS) inflict financial and emotional damage by defrauding victims under the guise of meaningful relationships. This research study examines RS trends in the U.S. through a quantitative analysis of web searches, news articles, research publications, and government reports from 2004 to 2023. This is the first study to use multiple sources for RS trend analysis. Results reveal increasing public interest and media coverage contrasted by a recent decrease in incidents reported to authorities. The frequency of research dedicated to RS has steadily grown but focuses predominantly on documenting the problem rather than developing solutions. Overall, findings suggest RS escalation despite declining official reports, which are likely obscured by low victim reporting rates. This highlights the need for greater awareness to encourage reporting enabling accurate data-driven policy responses. Additionally, more research must focus on techniques to counter these crimes. With improved awareness and prevention, along with responses informed by more accurate data, the rising RS threat can perhaps be mitigated.

## I. Introduction

Through the expansion of online social networking, romance scams (RS) are a particularly damaging type of social engineering attack which takes advantage of those seeking to forge new relationships. In this attack, victims receive convincing messages of love, care, and devotion which quickly establish strong and meaningful relationships [1]. Scammers exploit these relationships and extract financial resources from the victims until all money is taken or the relationships are terminated [2]. These attacks often cause financial, emotional, and psychological damage to the victims [3].

In the U.S., the organization that accumulates reports of suspected cybercrime "for investigative and intelligence purposes... and for public awareness" is the Internet Crime Complaint Center (IC3) which is run by the Federal Bureau of Investigation (FBI). In their most recent annual Internet Crime Report [4] released in March 2023, romance scams are shown to occur at a 22% lower frequency and with 23% lower dollar loss than the previous year. This suggests a significant change in direction after nearly a decade of continuous growth [4]. But evidence also suggests that RS is frequently not reported by victims [5]. Rather than focus on victim reports, this research study explores alternative methods of information gathering that may assist in gaining a more complete understanding of the pervasiveness of RS in society.

To achieve a greater understanding of RS trends, this study conducts a broad investigation using multiple sources. In doing so, this study seeks to answer the following research questions:

**RQ1** Are romance scams trending downward as IC3 suggests?
**RQ2** Can public information sources provide trend insights?
**RQ3** Where are RS researchers focusing their efforts?

These questions are answered through a quantitative analysis of web searches, news articles, research papers, government reports. The outcomes of this study intend to provide greater awareness of RS and encourage additional attention to the topic by individuals, organizations, and government agencies.

The rest of the paper is organized as follows: Section II outlines the methodology. Section III presents the results, followed by a discussion in Section IV. Section V touches on related work. Section VI presents the conclusions.

## II. Methodology

This study utilized a mixed methods approach to data collection using both quantitative and qualitative data which was converted to quantitative data for analysis. In data collection, the date range Jan 1, 2004 through Jan 31, 2023 was used to encompass all available data since the emergence of the modern RS. In addition, RS has gone by other names, thus the following synonymous keywords were used in searches: "romance scam", "sweetheart scam", "romance fraud", "online dating fraud", and "online dating scam". The following sections detail the sources and methods used for data collection, inclusion, exclusion, and categorization.

### A. Web Searches

To help answer RQ1 and RQ2, Google Trends (GT) was employed to analyze the frequency of the keyword searches on Google. Although GT has been identified as having data consistency concerns, it remains a viable data source [6]. For this study, the GT search is restricted to English and the U.S. and this results in the following search URL:

https://trends.google.com/trends/explore?date=all&geo=US
&q=romance%20scam,sweetheart%20scam,romance%20fraud,
online%20dating%20fraud,online%20dating%20scam&hl=en.

## B. News Articles

To further help answer RQ1 and RQ2, this study looked at news reports containing the RS keywords which indicate RS-related newsworthy events. Since most newsworthy events about criminal acts involve occasions when incidents are discovered or criminals are prosecuted, the frequency of these events over time can be a powerful indicator of the pervasiveness of RS in society.

Metadata from news articles was collected from Google using web scraping. The scraping process starts with the creation of numerous URLs which hold the data to be scraped. The following URL form was used:

https://www.google.com/search?q=%22[keyword]%22
&tbs=cdr:1,cd_min:[begin-date],cd_max:[end-
date]&tbm=nws

which generated 43,830 URLs. Web scraping software, Octoparse, processed each page and collected title, description, date, and link from each result. Duplicates and false positives (i.e., pages lacking keywords in the title or description) were removed, yielding 12,006 RS-related new articles. These articles were not reviewed individually.

## C. Research Publications

To help answer RQ1, RQ2 and RQ3, RS-related research publications were explored according to the following steps:

1) Literature was identified. Using both Google Scholar and ProQuest Database, searches were performed using the previously mentioned keywords.
2) Each piece of literature was assessed using inclusion criteria. For inclusion, RS was required to play a significant part in the content. This was a manual check.
3) Metadata was extracted and recorded from selected publications in a manner allowing quantitative analysis.
4) Each paper was assessed to determine whether it was wholly focused on RS or more generally focused on cybercrime with a significant portion related to RS. Areas of focus were also identified for each paper based on the researchers understanding of the content. A single paper may have multiple areas of focus.

## D. IC3 Internet Crime Reports

The FBI uses the IC3 to collect information on suspected cybercrimes. To help answer RQ1, all IC3 annual reports since 2003 were examined to determine the RS frequency and losses relative to other cybercrimes reported to the FBI [4]. From each report, the following information was gathered:

- the number of RS victims,
- the rank of the number of RS victims when compared to victims from other crimes,
- the value of the RS losses, and
- the rank of RS losses when compared to other crimes.

## E. U.S. Federal Trade Commission

To help answer RQ1, since 2018, as mandated by the Elder Abuse Prevention and Prosecution Act (EAPPA), the U.S.

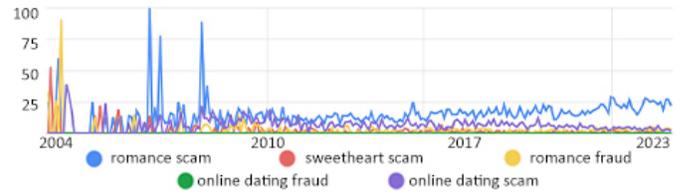

Fig. 1. Google Trends showing keyword interest over time.

Federal Trade Commission (FTC) has been tracking elderly fraud types including RS. Since the elderly are a primary target of RS, trends can be helpful in determining the trajectory of RS [7]. Data obtained from these reports includes the total reported losses and rank compared to other types of fraud [8].

## III. RESULTS

### A. Web Searches

When multiple keywords are used as they have been used in this study, the resulting data provides views of interest relative to the other keywords provided. Related to RQ1, Fig. 1 shows the following:

1) "Romance scam" is the most frequently searched term when compared with synonymous terms.
2) For each term, there was a spike or several spikes in interest soon after their introduction.
3) Since around 2010, the term "romance scam" shows a clear and steady upward trajectory while synonymous terms have fallen into disuse.
4) Searches are most often used to seek out information. This upward trajectory suggests there is an increasing desire by the U.S. population to learn more about romance scams.

These results which relate to RQ1 suggest that romance scams are trending upward.

### B. News Articles

The distribution of 12,006 RS-related news articles collected through webscraping, presented in Fig. 2, related to RQ1, shows the following:

1) Though there appears to be some leveling off between 2016 and 2018, overall, there has been a high rate of growth year-after-year since 2006.
2) Since 2008, the average annual growth rate is 66%.
3) The results give further confidence to answering RQ1 with a "no", romance scams are trending upward.
4) The upward trajectory suggests there will continue to be significant increases in the number of RS-related newsworthy events.

### C. Research Publications

Within the present research, 94 scholarly publications (listed in the Appendix at the end of the paper) were identified as containing a significant amount of RS-related content. The following details were found:

1) Related to RQ1 and RQ2, there has been a steady growth in the number of research papers which discuss RS.

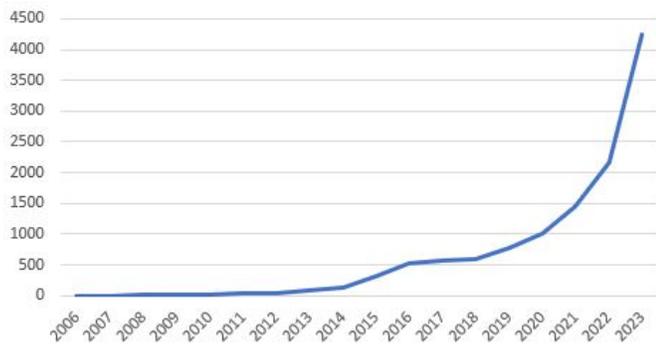

Fig. 2. Distribution of romance scams in news.

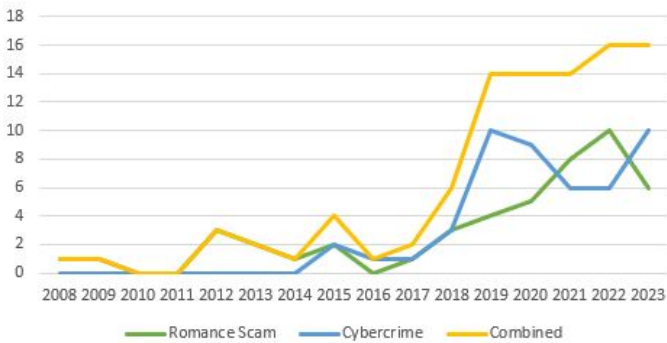

Fig. 3. Research publications by type.

2) Related to RQ3, there were 46 RS focused papers and 48 cybercrime focused papers which contained RS-related content. A distribution by year is presented in Fig. 3.
3) 2023 experienced a decline in research focused solely on RS with a tendency to expand the focus to include other cybercrimes.
4) Related to RQ3, as shown in Table I, the leading area of focus was on victim characteristics (29 instances) which focused on who was targeted or had the highest propensity to fall victim to RS.
5) The second highest area of focus covered the playbook (19 instances) which identified how the RS was enacted.
6) Additional areas of focus identified include older people (10), and reporting (10). Older people are often discussed because of their high propensity to be victimized. Reporting is a significant concern because this type of criminal act is seldom reported which may result in significant underestimations of the true damages.

### D. IC3 Internet Crime Reports

The topic of "Romance Fraud" appeared in the 2007 edition of the IC3 annual report, however no data was provided as it instead offered a warning of an emerging threat [4]. Data for "Romance Scams" first appeared in the 2011 edition of the report where romance scams were shown as being reported 5663 times and had the highest total loss of all criminal types on the report, $50,399,563.16 [4]. Related to RQ1, according to the IC3 Annual reports shown in Table II, the following details were found:

TABLE I
RS-RELATED RESEARCH FOCUS (PUBLICATIONS MAY HAVE MORE THAN ONE AREA OF FOCUS)

| Area of Focus | Total | Area of Focus | Total |
|---|---|---|---|
| Victim Characteristics | 29 | Law Enforcement | 2 |
| Playbook | 19 | Literature Review | 2 |
| Nigeria | 10 | Military | 2 |
| Older People | 10 | Online Dating | 2 |
| Perpetrators | 10 | Prevalence | 2 |
| Reporting | 10 | Social Engineering | 2 |
| Non-Financial Impacts | 9 | Deep Fakes | 1 |
| Blame | 5 | Deterrence | 1 |
| COVID-19 | 5 | Mass Marketing Fraud | 1 |
| Detection | 5 | Money Laundering | 1 |
| Fraudulent Profiles | 5 | Phishing | 1 |
| Legislation | 5 | Pig Butchering | 1 |
| Awareness | 4 | Sextortion | 1 |
| Linguistics | 4 | Treatment | 1 |
| Prevention | 4 | Victim Support | 1 |
| Identity Theft | 3 | Vigilantism | 1 |
| AI | 2 | | |

1) The victim rank, showing the number of victims compared to other cyber crimes, has always been low compared to the rank of reported losses indicating that victims of RS tended to lose more than victims of other cybercrimes.
2) There was a continuous increase in the number of RS victims and their losses from 2011 to 2021.
3) In 2022, there was a significant reduction in the number of victims, victim rank, total losses, and total rank.
4) From 2011 to 2020, RS losses ranked in the first or second place of all reported losses.
5) Starting in 2021 and continuing to 2022, RS loss rank started to decline.

TABLE II
IC3 INTERNET CRIME REPORTS [4]

| Year | Victims | V-Rank | Losses | L-Rank |
|---|---|---|---|---|
| 2022 | 19,021 | 11th | $ 735,882,192 | 5th |
| 2021 | 24,299 | 6th | $ 956,039,739 | 3rd |
| 2020 | 23,751 | 8th | $ 600,249,821 | 2nd |
| 2019 | 19,473 | 7th | $ 475,014,032 | 2nd |
| 2018 | 18,493 | 7th | $ 362,500,761 | 2nd |
| 2017 | 15,372 | 11th | $ 211,382,989 | 2nd |
| 2016 | 14,546 | 11th | $ 219,807,760 | 2nd |
| 2015 | 12,509 | 13th | $ 203,390,531 | 2nd |
| 2014 | 5,883 | 11th | $ 86,713,003 | 1st |
| 2013 | 6,412 | Not Available | $ 81,796,169 | 1st |
| 2012 | 4,476 | Not Available | $ 55,991,601 | 2nd |
| 2011 | 5,663 | Not Available | $ 50,399,563 | 1st |

TABLE III
FTC REPORTS OF REPORTED ELDERLY RS LOSSES [9]

| Year | Losses | Change | L-Rank |
|---|---|---|---|
| 2023 | 240M | +13% | 3rd |
| 2022 | 213M | +54% | 1st |
| 2021 | 139M | +66% | 1st |
| 2020 | 84M | +50% | 1st |
| 2019 | 56M | | 1st |

*E. U.S. Federal Trade Commission*

According to the FTC Annual reports shown in Table III, the following details which relate to RQ1 were found:

1) The losses reported to the FTC and included in these reports were separate from those reported to IC3.
2) Elderly losses to RS have continued to increase since records began in 2019, however, there has been a significant reduction in the growth rate from 2022 to 2023.
3) From 2019 to 2022, RS caused the greatest losses to victims. According to the 2023 FTC report, while RS losses increased, RS fell behind Investment Scams and Business Imposters as the most damaging fraud type affecting the elderly [9].
4) The FTC noted that the reported amounts are likely significantly lower than actual amounts because only 2% - 6.7% of elderly fraud is reported [9].

## IV. Discussion

Overall, in answering the research questions:

**RQ1** Combining the results across all information sources, romance scams appear to be generally trending upward.
**RQ2** There is a significant amount of public information that provides insights into RS trends including news articles, web searches, and scholarly works.
**RQ3** Within research literature related to romance scams, the most popular focus was "Victim characteristics" followed by "Playbook".

Perhaps one of the most disturbing findings was the low reporting rates identified by the FTC. Analysis of scholarly research showed that 10 of the selected reports focused on RS low reporting rates. The research identifies that RS leads to financial, emotional, and psychological harms which include the shame of falling for such scams [10]. Often, the victim is blamed by friends, family, and law enforcement. Additionally, when the scammer is from another country, law enforcement has little capability to do anything [10]. These factors contribute to the low reporting rate and support the FTC findings. Ultimately, the IC3 and FTC annual reports that are used to inform Congress, law enforcement, and the public of the state of cybercrimes are based on victim reporting which is significantly underreported.

Another discovery was the lack of research related to protecting individuals from RS. Over the years, there has been an evolution of the attack which has required scholars to alter their models and understanding of the RS playbook. Future developments are suspected to use AI and Deep fakes to enhance attacks and avoid detection [11]. And still, few methods exist to prevent these attacks. The most popular technique for prevention is awareness, but as Cross notes, the awareness messaging will need to be altered to address ever changing RS threats [11].

One of the most notable discoveries in this study is the seemingly exponential growth seen in news reports. These findings suggest that there should be nearly double the news reports on RS in 2024. Perhaps this is a sign of increased visibility, but it may also indicate increasing frequency. The sustained increases in web searches substantiate this claim as more people than ever are looking for RS information. Increased visibility may increase the reporting rates by ensuring victims know where and how to report the attacks. This may also increase the attention given to RS from researchers accelerating the development of mitigation methods.

## V. Related Work

There are three known systematic literature reviews that offer similar findings in the discovery of the increased frequency of scholarly work on RS [12]–[14]. These reviews provide analyses on areas of focus, and also include compiled knowledge including attacker motivation and methods, and scam detection techniques. [14] also suggests future research efforts should focus on technical countermeasures as these were found to be lacking. The IC3 reports provide the greatest similarity to this study in their determination of growth or decline over a period of 3 years [4]. None of the existing literature was found to employ alternative sources of data collection to establish RS trends.

## VI. Conclusion

This research study provides an examination of the growth of Romance Scams as seen through web searches, news reports, scholarly research, and U.S. Government reports by the IC3 and FTC. While web searches and scholarly research showed steady growth since the scam's early beginnings, news reports showed significant growth suggesting that RS will be increasingly found in the news in coming years. Though the FTC report showed limited increases, the IC3 report identified that RS are being reported less frequently. This may be due to a decrease in the number of attacks, or individuals increasingly not wanting to or not knowing how to report. This contradiction between collected public data and the IC3 report suggests that the IC3 report may not adequately represent RS in its findings and may need to revise methods to increase accuracy. Additionally, more research must focus on techniques to counter these socially engineered crimes. With an improvement in awareness and prevention, along with responses informed by more accurate data, the rising threat of RS can perhaps be mitigated.

APPENDIX

RS-related scholarly publications selected according to the methodology detailed in Section II C.